\documentstyle[12pt]{article}

\newcommand{\al}{\alpha}
\newcommand{\Al}{\mbox{\boldmath $\alpha$}}
\newcommand{\ba}{\begin{array}}
\newcommand{\be}{\begin{equation}}

\newcommand{\qc}[2]{{#1 \brack #2}}

\newcommand{\bea}{\begin{eqnarray}}

\newcommand{\ibea}{\begin{eqnarray}}

\newcommand{\non}{\nonumber}

\newcommand{\ea}{\end{array}}
\newcommand{\ee}{\end{equation}}
\newcommand{\eea}{\end{eqnarray}}

\newcommand{\lm}{\lambda}

\newcommand{\om}{\omega}
\newcommand{\epl}{\epsilon}

\newcommand{\rf}[1]{(\ref{eq:#1})}

\newtheorem{th}{Theorem}[section]
\newtheorem{prop}[th]{Proposition}
\newtheorem{df}[th]{Definition}

\begin{document}
\title{Colored Vertex Models, Colored IRF Models
and Invariants of Trivalent Colored Graphs}
\author{
Tetsuo Deguchi and
Yasuhiro Akutsu
$^{\dagger}$  }
\date{}
\maketitle
\begin{center}
\it
Department of Physics, Faculty of  Science,\\
     University of Tokyo, Hongo, Bunkyo-ku, Tokyo 113, Japan
\end{center}
\par \noindent
\begin{center}
\it
$^{\dagger}$
Department of Physics, Faculty of Science \\
Osaka University, Machikaneyama-cho 1-1, Toyonaka 560, Japan
\end{center}
{\bf Abstract}.
\par
We present  formulas for the Clebsch-Gordan coefficients
and the Racah coefficients for color representations
($N$-dimensional representations with
$q^{2N}=1$)
of $U_q(sl(2))$.
We discuss colored vertex models and
colored IRF (Interaction Round a Face) models from the
color representations of $U_q(sl(2))$.
We construct invariants of trivalent
colored oriented framed graphs
from color representations of $U_q(sl(2))$.

\newpage
\section{Introduction}
Recently  a new hierarchy
of $N$-state colored braid matrices
$(N=2,3, \cdots)$ has been constructed
which have color variables $\alpha$ and $\beta$.
 \cite{actsdeg1,degacts1,degacts2,degsn}
We regard colored braid as a braid on strings with colors.
{}From the colored braid matrices a new series
of isotopy invariants of colored links generalizing
the multivariable Alexander polynomial has been obtained. \cite{ado}
Using the limit $q^2 \rightarrow \omega$,
we have derived the $N$-state  colored braid matrices
from  infinite dimensional  representations of
$U_q(sl(2))$. \cite{inff}
Here $\omega$ is a primitive $N$-th root of unity.
Thus the colored braid matrices are derived from the $N$-dimensional
representations \cite{roche}
of  $U_q(sl(2))$ with $q^{2N}=1$.
We call the $N$-dimensional representation as (finite dimsional)
color representation. We  call infinite dimensional representation
with a free parameter infinite dimensional color representation.
We note that representation theory of
quantum groups with $q$ roots of unity
is given in Refs. \cite{concini,kac}.

In this paper we present formulas for
the Clebsch-Gordan coefficients
and the Racah coefficients of the color
representations of $U_q(sl(2))$.
We introduce colored IRF models and colored vertex models
from the color representations of
$U_q({\hat sl}(2))$.
We construct invariants of trivalent colored oriented graphs from the
Clebsch-Gordan coefficients of the color representations.
The new graph invariants give generalizations of
the multivariable Alexander polynomial of links.

%
%
%
%

There are various viewpoints associated with the colored vertex models.
The $N=2$ case of the colored vertex model
corresponds to  the trigonometric limit
of the Felderhof's solution
\cite{felderhof}
of the free fermion model. \cite{fanwu}
The non-colored case
of the trigonometric limit is related to the
Lie superalgebra $gl(1|1)$ ($sl(1|1)$). \cite{graded}
\cite{schultz,perk,kulskl}
%
%
It has been pointed out
\cite{rozansky} that the colored case of the trigonometric limit
is related to $U_q(gl(1|1))$. From the different viewpoint
it has also been shown \cite{murakami2}
that the colored case of the trigonometric limit is
related to $U_q(sl(2))$.

The free fermion model and the colored vertex models are related
to the link polynomials, which vanish for disconnected links.
{}From the $gl(m|m)$ model we have the Alexander polynomial.
\cite{graded}
A state model for the multivariable Alexander polynomial
was constructed from the trigonometric limit of the free fermion model.
\cite{murakami}
Several examples of braid matrices with the vanishing property
were calculated by Lee, Couture
and Schmeing, by directly solving the braid relation up to the 6 state case.
\cite{lee}

%
%
%
%

The outline of this paper is given in the following.
In \S 2 we give the Clebsch-Gordan coefficents and the Racah coefficients
of infinite dimensional representations of $U(sl(2))$.
In \S 3 we present formulas
for the Clebsch-Gordan coefficients
and the Racah coefficients
for infinite-dimensional  color
representations of $U_q(sl(2))$.
By taking the limit $q^2 \rightarrow \omega$  we derive formulas  for
 the Clebsch-Gordan coefficients and the Racah coefficients
of finite-dimensional color  representations of $U_q(sl(2))$.
In \S 4 we discuss the colored vertex models from the viewpoint of
color representations
of the $U_q({\hat sl(2)})$.
In \S 5 we construct IRF models
associated with color representations of
 $U_q(sl(2))$, which we have called colored IRF models.
In \S 6 we obtain new invariants of trivalent colored graphs
from the Clebsch-Gordan coefficients of color representations.

\setcounter{equation}{0}
\renewcommand{\theequation}{2.\arabic{equation}}
\section{ CGC
of infinte dimensional representations of $U(sl(2))$}

Let us discuss representations
of the universal enveloping algebra $U(sl(2))$.
The defining relations of the algebra $U(sl(2))$
are given by the following.
\bea
\mbox{[}H, X^{\pm} \mbox{]} &= & \pm 2 X^{\pm},  \non \\
\mbox{[} X^{+}, X^{-} \mbox{]} & = & H .
\eea
The comultiplication is given by
\bea
\Delta(X^{\pm})
& = & X^{\pm} \otimes I + I \otimes X^{\pm} , \non \\
\Delta(H)
& = & H \otimes I + I \otimes H .
\eea
Let us define infinite dimensional representation
$(\pi^{p}, V^{(p)})$ of $sl(2)$.
Let $V^{(p)}$ be an infinite dimensional vector space
over ${\bf C}$ with basis
$e_0, e_1, \cdots$, where $e_a$ is a basis vector in $V^{(p)}$
with the property $(e_a)_{b} = \delta_{b a}$.
We define the matrix elements of the generators as follows.
\bea
 \left( \pi^{p} (X^{+}) \right)^a_{b}
 &=& {\sqrt{(2p - a)(a+1)}} \cdot \delta_{a+1, b} ,  \non \\
\left( \pi^{p} (X^{-}) \right)^a_b &=&
{\sqrt{a(2p-a+1)}} \cdot \delta_{a-1, b} , \non \\
\left( \pi^{p} (H)  \right)^a_{b}
&=& (2p - 2a) \cdot \delta_{a, b}.  \label{eq:infrep}
\eea
The variable $a$ takes a nonnegative integer;
$a=0,1,\cdots$.
The basis vectors for the infinite dimensional representation
are labeled by $(p, z)$, where $p \in {\bf C}$ and
$z \in {\bf Z}_{\ge 0}$.
Here the symbol ${\bf Z}_{\ge 0}$ denotes
the set of nonnegative integers.
We write the representation $(\pi^{p}, V^{(p)})$
simply by $V^{(p)}$, and
the basis vector $e_z$ by $|p,z>$.
In terms of the basis vector the representation \rf{infrep} is
expressed as follows.
\bea
X^+ |p,z> & = & {\sqrt{(2p-z+1)z}} |p, z-1>,  \non \\
X^- |p,z> & = & {\sqrt{(2p-z)(z+1)}} |p, z+1>,  \non \\
H |p,z> & = & (2p-2z) |p, z> . \label{eq:kets}
\eea

Let us discuss decomposition of the tensor product
$V^{(p_1)} \otimes V^{(p_2)}$.
We assume the following
\be
|p_1,p_2; p_3,z_3> = \sum_{z_1,z_2}
C(p_1,p_2,p_3; z_1,z_2,z_3)
|p_1, z_1> \otimes |p_2, z_2> .  \label{eq:defcg}
\ee
{}From the comultiplication rule
$\Delta(H) = H \otimes I + I \otimes H$ ($H_{tot} = H_1 + H_2$)
we can show
\be
C(p_1,p_2,p_3; z_1,z_2,z_3) = 0, \mbox{ unless}
\quad p_1-z_1+p_2-z_2 = p_3-z_3. \label{eq:charge}
\ee
The relation \rf{charge} gives the charge conservation law.
Using  \rf{charge} and  the condition $z_i \in {\bf Z}_{\ge 0}$,
we can show
\be
C(p_1,p_2,p_3; z_1,z_2,z_3) = 0, \mbox{ unless} \quad
p_3=p_1+p_2-n, n \in {\bf Z}_{\ge 0}.
\label{eq:addition}
\ee
Thus we have
for $n \in {\bf Z}_{\geq 0}$
\bea
&& |p_1,p_2; p_1+p_2-n ,z_3>  \non \\
& = & \sum_{z_1=0}^{z_3+n}
C(p_1,p_2,p_1+p_2-n;z_1, z_3+n-z_1, z_3)
|p_1, z_1> \otimes |p_2, z_3+n-z_1> .  \non \\
\label{eq:br}
\eea
The fusion rules for the tensor product
$V^{(p_1)} \otimes V^{(p_2)}$
are summarized
as follows.
\be
V^{(p_1)} \otimes V^{(p_2)}
= \sum_{p_3} N_{p_1 p_2}^{p_3} V^{(p_3)}
\ee
where
\bea
N_{p_1 p_2}^{p_3} & = & 1 \mbox{ for }
p_3=p_1+p_2 - n , n \in {\bf Z}_{\ge 0} , \non \\
& = & 0, \quad \mbox{ otherwise} .
\eea
We assume the following convention of the phase factors
of the CG coefficients.
\be
C(p_1,p_2,p_1+p_2-n; 0,n,0) = 1.
\ee
Then the Clebsch-Gordan coefficients
for infinite dimensional representations
are given by the following.
\bea
&& C(p_1, p_2, p_1+p_2-n; z_1, z_2, z_3)
= \delta(z_3, z_1+z_2-n) \non \\
&& \times {\sqrt{(2p_1+2p_2-2n+1)}}
\left(
{\frac {(2p_1-n)!(2p_2-n)!(n)!} {(2p_1+2p_2-n+1)!} } \right)^{1/2}
 \non \\
&& \times {\sqrt{(2p_1-z_1) !  (2p_2-z_2) ! (2p_1+2p_2-2n-z_3) ! (z_1)! (z_2)!
 (z_3)!}} \non \\
&& \times \sum_{\nu} {\frac {(-1)^{\nu}} {(\nu)! (n-\nu)! }}
 {\frac 1 {(z_1-\nu)! (z_2-n+\nu)! }}  \non \\
&& \times
{\frac 1 {(2p_1 -n -z_1 + \nu)! (2p_2 -z_2 - \nu)!}} . \label{eq:ppp}
\eea
Here we have defined the symbol $(p)!$ through the Gamma function
\be
(p)! = \Gamma(p+1). \label{eq:gamma}
\ee
The sum over the integer $\nu$
in \rf{ppp} is taken
under the following condition
\be
\mbox{max}\{0, n-z_2 \}  \le \nu  \le \mbox{min}\{n, z_1 \}.
\ee
We can derive the Clebsch-Gordan coefficients \rf{ppp}
following Racah's derivation \cite{rose} (see also Appendix A).

Hereafter we assume that the symbols $p$ and $p_j$ denote
 complex parameters ($p$, $p_j \in {\bf C}$), and
$2j$ and $2j_k$ denote nonnegative integers
($2j, 2j_k \in {\bf Z}_{\ge 0}$).
We also assume that
$2\mu_j$ ($j=1,2,3$) denote either  non-negative integers
($2\mu_j \in {\bf Z}_{\ge 0}$) or complex parameters ($\mu_j \in {\bf C}$).

Let $V^{(j)}$ denote the
 representation with dimension $2j+1$ (spin $j$ representation).
Let us consider the following decomposition of the tensor product.
\be
V^{(\mu_1)} \otimes V^{(\mu_2)}
= \sum_{p_3} N_{\mu_1 \mu_2}^{p_3} V^{(p_3)} , p_3 \in {\bf C}.
\ee
Here we consider the two cases,
(1)
$(\mu_1,\mu_2) =(p_1, j_2)$
($p_1 \in {\bf C}$ , $2j_2 \in$ ${\bf Z}_{\ge 0}$);
(2)
$(\mu_1,\mu_2) = (j_1, p_2)$
($2j_1 \in{\bf Z}_{\ge 0}$, $p_2 \in {\bf C}$).

Then we have the following fusion rules.
\par \noindent
Case (1) $(\mu_1,\mu_2) =(p_1, j_2)$
($p_1 \in {\bf C}$ , $2j_2 \in$ ${\bf Z}_{\ge 0}$);
\bea
N_{p_1 j_2}^{p_3} & = & 1 \mbox{ if }
p_3=p_1+j_2 - n , 0 \le n \le 2j_2,
n \in {\bf Z}_{\ge 0} , \non \\
& = & 0, \quad \mbox{ otherwise} .
\eea
\par \noindent
Case (2)  $(\mu_1,\mu_2) = (j_1, p_2)$
($2j_1 \in{\bf Z}_{\ge 0}$, $p_2 \in {\bf C}$)
\bea
N_{j_1 p_2}^{p_3} & = & 1 \mbox{ if }
p_3=j_1+p_2 - n , 0 \le n \le 2j_1,  n \in {\bf Z}_{\ge 0} , \non \\
& = & 0, \quad \mbox{ otherwise} . \label{eq:brpjp}
\eea


The Clebsch-Gordan coefficients are given by the following.
\bea
&& C(\mu_1, \mu_2, \mu_1+\mu_2-n; z_1, z_2, z_3)
= \delta(z_3, z_1+z_2-n) \non \\
&& \times {\sqrt{(2\mu_1+2\mu_2-2n+1)}}
\left( {\frac {(\mu_1-n)!(2\mu_2-n)!(n)!} {(2\mu_1+2\mu_2-n+1)!} }
\right)^{1/2}
 \non \\
&& \times {\sqrt{(2\mu_1-z_1) !  (2\mu_2-z_2) ! (2\mu_1+2\mu_2-2n-z_3) !
(z_1)! (z_2)!  (z_3)!}} \non \\
&& \times \sum_{\nu} {\frac {(-1)^{\nu}} {(\nu)! (n-\nu)! }}
 {\frac 1 {(z_1-\nu)! (z_2-n+\nu)! }} \non \\
&& \quad \times {\frac 1 {(2\mu_1 -n -z_1 + \nu)! (2\mu_2 -z_2 - \nu)!}}
 \label{eq:pjp}
\eea
Here the sum over the integer $\nu$ is taken  under the following conditions:
\par \noindent
Case (1)  $\mu_1= p_1, \mu_2 = j_2$,
\be
\mbox{max}\{0, n-z_2 \}  \le \nu  \le \mbox{min}\{n, z_1, 2j_2-z_2 \}.
\ee
\par \noindent
Case (2) $\mu_1 = j_1, \mu_2 = p_2$,
\be
\mbox{max}\{0, n-z_2, n+z_1-2j_1 \}  \le \nu  \le \mbox{min} \{n, z_1 \}.
\ee

We note that the formula \rf{pjp} can be derived from \rf{ppp}
by taking the limit $p_2 \rightarrow j_2$.
The branching  rule \rf{brpjp}
can be shown through this limiting procedure.

\par
The Clebsch-Gordan coefficients satisfy the orthogonality relations
for the cases $(\mu_1, \mu_2)=(p_1,p_2), (p_1, j_2)$,
and $(j_1,p_2)$.
\bea
\delta_{nn^{'}}
& = & \sum_{z_1} C(\mu_1, \mu_2, \mu_1+\mu_2-n ; z_1, \tau -z_1, \tau - n) \non
\\
&& \quad \times C(\mu_1, \mu_2, \mu_1 + \mu_2 - n^{'} ;
 z_1, \tau-z_1, \tau-n^{'}) .
\label{eq:oppp}
%
\eea

Let us discuss the Racah coefficients for finite and infinite dimensional
representations. Using the formulas
for the Clebsch-Gordan coefficients \rf{ppp}, and \rf{pjp}
we can calculate the Racah coefficients.
We recall that  $2\mu_i$ $(i=1,2,3)$ is either a nonnegative integer
$(2\mu_i \in {\bf Z}_{\ge 0}$) or
a complex parameter ($\mu_i \in {\bf C}$).
We introduce $\mu, \mu^{'}, \mu^{''}$ by
\be
\mu =\mu_1+\mu_2+\mu_3-n, \mu^{'}=\mu_1+\mu_2-n^{'},
\mu^{''} = \mu_2+\mu_3-n^{''} ,
\ee
where $n, n^{'}, n^{''} \in {\bf Z}_{\ge 0}$.

The Racah coefficients
$W(\mu_1,\mu_2,\mu,\mu_3;\mu^{'},\mu^{''})$
are defined by \cite{rose}
\bea
&& |\mu_1\mu_2(\mu^{'})\mu_3\mu z> = \non \\
&& = \sum_{\mu^{''}} {\sqrt{(2\mu^{'}+1)(2\mu^{''}+1)}}
W(\mu_1,\mu_2,\mu,\mu_3;\mu^{'},\mu^{''})
|\mu_1,\mu_2\mu_3(\mu^{''})\mu z> . \non \\
&& \label{eq:defrc}
\eea
We note that the Racah coeffcients do not depend on
$z$ by the definition \rf{defrc}.
{}From the definition of the Racah coefficients, we have
\bea
&& C(\mu_1,\mu_2, \mu^{'}; w_1, w_2, w_1+w_2-n^{'})
C(\mu^{'}, \mu_3, \mu; w_1+w_2-n^{'}, w_3, w_1+w_2+w_3-n)  \non \\
& = & \sum_{\mu^{''}} R_{\mu^{''} \mu^{'}}
C(\mu_1,\mu^{''}, \mu; w_1, w_2+w_3-n^{''}, w_1+w_2+w_3 -n) \non \\
&& \times C(\mu_2, \mu_3, \mu^{''}; w_2, w_3, w_2+w_3-n^{''} ) .
\label{eq:rcc}
\eea
Here $w_1,w_2, w_3,  \in {\bf Z}_{\ge 0}$.
Applying the orthogonality relation \rf{oppp}
to \rf{rcc}
we have a formula for
the Racah coefficients.
The derivation of the Racah coefficients consists of the following
procedures.
(1) Let us set $w_2+w_3=z^{''}$.
(2) Then multiply both the left hand side and right hand side
of \rf{rcc}
by $\sum_{w_2} C(\mu_2,\mu_3, \mu_2+\mu_3-N^{''};
w_2, z^{''}-w_2, z^{''}-N^{''})$ .
(3) Rewrite $N^{''}$ by $n^{''}$, and introduce
$z=w_1+z^{''}=w_1+w_2+w_3$.
(4) Multiply both sides
by $\sum_{w_1} C(\mu_1, \mu_2+\mu_3-n^{''}, \mu_1+\mu_2+\mu_3-n ;
w_1, z-w_1, z+n^{''}-n)$.

Thus we have the Racah coefficients
$W(\mu_1, \mu_2, \mu, \mu_3 ; \mu^{'}, \mu^{''})$ as follows.
\bea
R_{\mu^{''} \mu^{'}}& = &{\sqrt{(2\mu^{'}+1)(2\mu^{''}+1)}}
W(\mu_1, \mu_2, \mu, \mu_3 ; \mu^{'}, \mu^{''}) \non \\
&= & \sum_{w_1}\sum_{w_2}
C(\mu_1, \mu_2, \mu^{'}; w_1, w_2, w_1+w_2-n^{'}) \non \\
&& \times C(\mu^{'}, \mu_3, \mu; w_1+w_2-n^{'}, z-w_1-w_2, z-n) \non \\
&& \times C(\mu_2, \mu_3, \mu_2+\mu_3-n^{''} ; w_2, z-w_1-w_2, z-w_1-n^{''})
\non \\
&& \times C(\mu_1, \mu^{''}, \mu; w_1, z-w_1, z+n^{''}-n ) .
\eea

\setcounter{equation}{0}
\renewcommand{\theequation}{3.\arabic{equation}}
\section{ CGC of color representations of $U_q(sl(2))$}
\subsection{Infinite dimensional case }
The generators of the
algebra are $\{X^{+}, X^{-}, H \}$ with
the defining relations \cite{drinfeld,jimbo1}
\bea
\mbox{[} H, X^{\pm} \mbox{]} & = & \pm 2 X^{\pm} , \non \\
\mbox{[} X^{+}, X^{-} \mbox{]} & = & {\frac {q^{H}-q^{-H}}
{q-q^{-1}} } .
\eea
The comultiplication  is given by
\bea
\Delta(H) & = & H \otimes I + I \otimes H , \non \\
\Delta(X^{\pm}) & = & X^{\pm} \otimes q^{H/2} + q^{-H/2} \otimes X^{\pm} .
\eea
We use the following symbols for $q$-analogs.
\be
[n]_q  = {\frac {q^{n} - q^{-n}} {q - q^{-1}} } , \quad
[n]_q ! = \prod_{k=1}^{n} [k]_q , \quad
[p;n]_q ! = \prod_{k=0}^{n-1} [p-k]_q .
\ee
Here $n$ is a positive integer.
For the case $n=0$ we assume
\be
[0]_q!  =  [p;0]_q ! = 1 .
\ee

\par
Let us define an infinite dimensional representation
$(\pi^{p}_q, V^{(p)})$ of $U_q(sl(2))$.
We recall that $p$ is a complex parameter.
Let $V^{(p)}$ be an infinite dimensional vector space
over ${\bf C}$ with basis
$e_0, e_1, \cdots$, where $e_a$ is a basis vector in $V^{(p)}$
with the property $(e_a)_{b} = \delta_{b a}$.
We define matrix elements of the representations of the generators
$X^{+}$, $X^{-}$ and $K=q^H$ for  $(\pi^{p}_q, V^{(p)})$ as follows.
\bea
 \left( \pi^{p}( X^{+})_q \right)^a_{b}
 &=& {\sqrt{[2p - a]_q[a+1]_q}} \cdot \delta_{a+1, b} ,  \non \\
\left( \pi^{p}( X^{-})_q \right)^a_b &=&
{\sqrt{[2p-a+1]_q[a]_q}} \delta_{a-1,b}, \non \\
\left( \pi^{p}( H )_q \right)^a_{b}
&=& (2p - 2a) \cdot \delta_{a, b}
 .    \label{eq:colrep}
\eea
Here $a$,$b$ are nonnegative integers $(a,b=0,1,\cdots)$,
 and $p$ $\in {\bf C}$.
It is easy to see that the operators defined in  \rf{colrep}
satisfy the defining relations of
the algebra $U_q(sl(2))$ with $q$ generic.
We call the representation in  \rf{colrep} infinite dimensional
color representation.
We write the basis vector $e_a(p)$ by $|p, z>_q$,
 and the representation
 $(\pi^{p}_q, V^{(p)})$ by $V^{(p)}_q$.

Let us discuss decomposition of the tensor product
$V_q^{p_1} \otimes V_q^{(p_2)}$.
We define the Clebsch-Gordan coefficients by the following
\be
|p_1,p_2; p_3,z_3>_q = \sum_{z_1,z_2}
C(p_1,p_2,p_3; z_1,z_2,z_3)_q
|p_1, z_1>_q \otimes |p_2, z_2>_q .  \label{eq:dfqppp}
\ee
{}From the comultiplication rule
we can show
\be
C(p_1,p_2,p_3;z_1,z_2,z_3)_q = 0, \mbox{ unless} \quad
p_1-z_1+p_2-z_2 = p_3-z_3.
\label{eq:qcharge}
\ee
The relation \rf{qcharge} leads to  the charge conservation law.
Using  \rf{qcharge} and  the condition $z_i \in {\bf Z}_{\ge 0}$,
we can show
\be
C(p_1,p_2,p_3; z_1,z_2,z_3)_q = 0, \mbox{ unless } \quad
p_3=p_1+p_2 -n, n \in {\bf Z}_{\ge 0}.  \non \\
\label{eq:qaddition}
\ee
Thus we have following.
\bea
&& |p_1,p_2; p_1+p_2-n ,z_3>_q \non \\
& = & \sum_{z_1=0}^{z_3+n}
C(p_1,p_2,p_1+p_2-n; z_1,z_3+n-z_1, z_3)_q
|p_1, z_1>_q \otimes |p_2, z_3+n-z_1>_q .  \non \\
&& \qquad \label{eq:qbr}
\eea
Here $n \in {\bf Z}_{\geq 0}$.
The Clebsch-Gordan coefficients
for infinite dimensional representations are
given by the following.
\bea
&& C(p_1, p_2, p_1+p_2-n; z_1, z_2, z_3)_q
= \delta(z_3, z_1+z_2-n) \non \\
&& \times {\sqrt{[2p_1+2p_2-2n+1]_q}}
{\sqrt{[n]_q! [z_1]_q!  [z_2]_q!  [z_3]_q!}}
\non \\
&& \times q^{(n-n^2)/2 + (n-z_2)p_1 +(n+z_1)p_2}
\non \\
&& \times \sum_{\nu} {\frac {(-1)^{\nu} q^{-\nu(2p_1+2p_2-n+1)} }
{[\nu]_q! [n-\nu]_q! [z_1-\nu]_q! [z_2-n+\nu]_q!} } \non \\
&& \times {\sqrt{\frac
{[2p_1-n; z_1-\nu]_q ! [2p_1-z_1; n-\nu]_q!
 [2p_2-n; z_2+\nu -n]_q ! [2p_2-z_2; \nu]_q ! }
 {[2p_1 +2p_2 -n+1; z_1+z_2+1]_q! }}} . \non \\
 \label{eq:qppp}
\eea
Here $p_3=p_1+p_2-n, n \in {\bf Z}_{\ge 0}$, and
the sum over  the integer $\nu$
in \rf{qppp} is taken
under the following condition
\be
\mbox{max}\{0, n-z_2 \}  \le \nu  \le \mbox{min}\{n, z_1 \}.
\ee
We shall derive the symmetric expression \rf{qppp} of the Clebsch-Gordan
coefficients \rf{qppp} in Appendix A.

Let $V^{(j)}_q$ be the spin $j$ representation of $U_q(sl(2))$.
Let us discuss decomposition of the tensor product
$V_q^{(p_1)} \otimes V_q^{(j_2)}$ and
$V_q^{(j_1)} \otimes  V_q^{(p_2)}$.
It is easy to see the following rules.
\bea
&& C(\mu_1,\mu_2,p_3;z_1,z_2,z_3)_q  =  0, \non \\
&& \mbox{ unless} \quad \mu_1-z_1+\mu_2-z_2 = p_3-z_3,
 p_3=\mu_1+\mu_2 -n \quad ( n \in {\bf Z}_{\ge 0}), \non \\
&& \quad \mbox{ and } \left\{
\begin{array}{c}
0 \le n \le 2j_2,
0 \le z_2 \le 2j_2,
  \mbox{ if } (\mu_1, \mu_2)= (p_1 ,  j_2), \non \\
				0 \le n \le 2j_1,
        0 \le z_1 \le 2j_1,
				\mbox{ if }  (\mu_1, \mu_2)= (j_1 ,  p_2).
\end{array} \right. \\
\eea

The Clebsch-Gordan coefficients are given by
\bea
&& C(\mu_1, \mu_2, \mu_1+\mu_2-n; z_1, z_2, z_3)_q
= \delta(z_3, z_1+z_2-n) \non \\
&& \quad \times {\sqrt{[2\mu_1+2\mu_2-2n+1]_q}}
{\sqrt{[n]_q! [z_1]_q!  [z_2]_q!  [z_3]_q!}}
\non \\
&& \quad \times q^{(n-n^2)/2 +(n-z_2)\mu_1 + (n+z_1)\mu_2}  \non \\
&& \quad \times \sum_{\nu} {\frac {(-1)^{\nu}
q^{-\nu(2\mu_1+2\mu_2-n+1)} }
{[\nu]_q! [n-\nu]_q! [z_1-\nu]_q! [z_2-n+\nu]_q!}} \non \\
&& \times
\left( {\frac{[2\mu_1-n; z_1-\nu]_q ! [2\mu_1-z_1; n-\nu]_q!
[2\mu_2-n; z_2+\nu -n]_q ! [2\mu_2-z_2; \nu]_q ! }
{[2\mu_1 +2\mu_2 -n+1; z_1+z_2+1]_q! }} \right)^{1/2} .
\non \\
&& \label{eq:qpjp}
\eea
Here the sum over  the integer $\nu$
in \rf{qpjp} is taken
under the following conditions.
\par \noindent
Case (1): $\mu_1 = p_1, \mu_2 = j_2$
\be
\mbox{max}\{0, n-z_2 \}  \le \nu  \le \mbox{min}\{n, z_1, 2j_2-z_2 \}.
\ee
\par \noindent
Case (2): $\mu_1=j_1, \mu_2=p_2$
\be
\mbox{max}\{0, n-z_2, n+z_1-2j_1 \}  \le \nu  \le \mbox{min}\{n, z_1 \}.
\ee

For convenience we give
the Clebsch-Gordan coefficients for finite dimensional representations
$V^{(j_i)}, (i=1,2,3)$  in the following
\cite{kr} (see also \cite{nomura1}).
\bea
&& C(j_1, j_2, j_3; z_1, z_2, z_3)_q  = \delta(z_3, z_1+z_2-n) \non \\
&& \times ([2j_3+1]_q)^{1/2}  \Delta_q(j_1j_2j_3) \non \\
&& \times q^{j_1(j_1+1) + j_2(j_2+1)-j_3(j_3+1)
+2(j_1j_2 + j_1(j_2 -z_2) - j_2(j_1-z_1))/2} \non \\
&& \times ([2j_1-z_1]_q ! [z_1]_q! [2j_2-z_2]_q ! [z_2]_q!
[2j_3-z_3]_q ! [z_3]_q! )^{1/2} \non \\
&& \times \sum_{\nu} {\frac {(-1)^{\nu}} {[\nu]_q!}}
 {\frac {q^{-\nu(j_1+j_2+j_3+1)}}
 {[j_1+j_2-j_3-\nu]_q! [z_1-\nu]_q!}}  \non \\
&& \times {\frac 1 {[2j_2 -z_2 -\nu]_q! [j_3 + j_1-j_2 -z_1 +\nu]_q!
[ j_3-j_1 - j_2+ z_2 +\nu]_q!}} . \non \\
&& \label{eq:qjjj}
\eea
Here the sum over the integer $\nu$ is such that
\bea
&& \mbox{max} \{ 0, j_2-j_3-j_1+z_1, j_1+j_2-j_3-z_2 \}
\le \nu \le \mbox{min} \{ z_1, j_1+j_2-j_3, 2j_2-z_2 \} .
\non \\
&&
\eea
The symbol $\Delta_q(abc)$ has been defined  by
\be
\Delta_q(abc) =
{(\frac {[a+b-c]_q! [c+a-b]_q! [b+c-a]_q!} {[a+b+c+1]_q!}})^{1/2} .
\ee
If we replace   $p_i$ by $j_i$ ($i=1,2,3$),
then the formal expression of the Clebsch-Gordan coefficients
for infinite dimensional
representations is consistent with
the finite dimensional one \rf{qjjj}.

The Clebsch-Gordan coefficients satisfy the orthogonality relations
for the cases $(\mu_1, \mu_2)=(p_1,p_2), (p_1, j_2)$, and $(j_1,p_2)$.
\bea
 \delta_{nn^{'}}
&= & \sum_{z_1} C(\mu_1, \mu_2, \mu_1+\mu_2-n;
 z_1, \tau -z_1, \tau - n)_q \non \\
&& \quad \times C(\mu_1, \mu_2, \mu_1+\mu_2-n^{'};
z_1, \tau-z_1, \tau-n^{'})_q ,
\label{eq:oqppp}
\eea

We can calculate the Racah coefficients
using  the Clebsch-Gordan coefficients and the orthogonality
relations \rf{oqppp} as in \S 2.
The Racah coefficients are given by
\bea
R_{\mu^{''} \mu^{'}}(q) & = &{\sqrt{[2\mu^{''}+1]_q [2\mu^{'}+1]_q}}
W(\mu_1,\mu_2,\mu,\mu_3;\mu^{'},\mu^{''})_q \non \\
&= & \sum_{w_1}\sum_{w_2} C(\mu_1, \mu_2, \mu^{'}; w_1, w_2, w_1+w_2-n^{'})_q
\non \\
&& \times
C(\mu^{'}, \mu_3, \mu; w_1+w_2-n^{'}, z^{''}-w_2, w_1+z^{''}-n)_q \non \\
&& \times
C(\mu_2, \mu_3, \mu_2+\mu_3-n^{''} ; w_2, z^{''}-w_2, z^{''}-n^{''})_q
\non \\
&& \times C(\mu_1, \mu^{''}, \mu; w_1, z^{''}, w_1+z^{''}+n^{''}-n )_q .
\eea
Here we have assumed that
$2\mu_i (i=1,2,3)$ is either a nonnegative integer
($2 \mu_i \in {\bf Z}_{\ge 0}$)
or a complex parameter ($\mu_i \in {\bf C}$).
We have defined $\mu, \mu^{'}, \mu^{''}$ by
\be
\mu =\mu_1+\mu_2+\mu_3-n, \mu^{'}
=\mu_1+\mu_2-n^{'}, \mu^{''} = \mu_2+\mu_3-n^{''} .
\ee
where $n, n^{'}, n^{''} \in {\bf Z}_{\ge 0}$.


\subsection{Finite dimensional cases }

Let  $\omega$ be a primitive $N$-th root of unity:
\be
\om= \exp( {\frac {2\pi i s } N} ), \qquad (N,s) =1.
\ee
Here the symbol $(a,b)=1$ denotes that the integers $a$ and $b$
have no common divisor except 1.
We write by $\epsilon$ a square root of $\omega$:
$\epsilon= \om^{1/2}= \exp({\pi i s}/N), \quad (N,s)=1$.

We take the limit $q \rightarrow \epl$
in the infinite dimensional representation  \rf{colrep}.
Then we have
\be
\lim_{q \rightarrow \epl} \left( \pi^{p}( X^{-})_q
\right)^{N}_{N-1} = 0 . \label{eq:vanish}
\ee
{}From the property  \rf{vanish}
we can restrict the infinite dimensions into $N$ dimensions:
$V^{(p)}_q(\infty) \rightarrow V^{(p)}_{\epl}(N)$,
where $V^{(p)}_{\epl}(N)$ is an $N$-dimensional
vector space with basis $e_0, \cdots, e_{N-1}$.
For $a,b =0,1, \cdots N-1$ we have the following matrix representations
\bea
\left( \pi^{p}(X^{+})_{q=\epl} \right)^a_{b} &=&
\left( [2p - a]_{\epl} [a+1]_{\epl} \right)^{1/2}
\cdot \delta_{a+1, b},  \non \\
\left( \pi^{p}(X^{-})_{q =\epl} \right)^a_{b} &=&
\left(
[2p-a+1]_{\epl} [a]_{\epl} \right)^{1/2}
\cdot \delta_{a-1, b}, \non \\
\left( \pi^{p}(K)_{q=\epl}
\right)_{ab} &=& \epl^{(2p - 2a)} \cdot \delta_{a, b} .
 \label{eq:restrict}
\eea
We write the finite dimensional color representation
$\{ \pi^p_{\epl}, V^{(p)} \}$ by $V^{(p)}_{\epl}$.
We write the basis vector $e_a$ by $|p,z>_{\epl}$.
The basis vectors for $V^{(p)}_{\epl}$ are
 $\{|p,z>_{\epl}, z=0, \cdots, N-1\}$.

Let us discuss decomposition of the tensor product
$V_{\epl}^{(p_1)} \otimes V_{\epl}^{(p_2)}$.
We take the limit
 $q \rightarrow \epl$ in the expression of the Clebsch-Gordan
 coefficients \rf{qppp}.
 Since there is no singularity in
 the limiting process
in the expression of the Clebsch-Gordan coefficients,
we have the fusion rule in the following.
\be
V^{(p_1)}_{\epl} \otimes V^{(p_2)}_{\epl}
= \sum_{p_3} N_{p_1,p_2}^{p_3} V^{(p_3)}_{\epl},
\ee
where
\bea
N_{p_1,p_2}^{p_3} & = & 1 ,
\mbox{ for } p_3=p_1+p_2, p_1+p_2-1, \cdots, p_1+p_2-N+1. \non \\
& = & 0, \mbox{ otherwise} .  \label{eq:freppp}
\eea
We note that
the condition $0 \le n \le N-1$ is derived from
the factor ${\sqrt{[n]_q!}}$ in \rf{qppp}.
In terms of the Clebsch-Gordan coefficients we have
\bea
&& C(p_1,p_2,p_3;z_1,z_2,z_3)_{\epl}  =  0, \non \\
&& \mbox{ unless} \quad
p_1-z_1+p_2-z_2 = p_3-z_3,
  p_3=p_1+p_2 -n, \non \\
&& \quad  0 \le z_i \le N-1 \mbox{ for } i=1,2,3,
 \mbox{ and } 0 \le n \le N-1,
n \in {\bf Z}_{\ge 0}. \non \\
\label{eq:breppp}
\eea

By taking the limit $q \rightarrow \epl$ in  \rf{qppp}
we obtain the Clebsch-Gordan coefficients
for finite dimensional color representations.
Let us set $p_3=p_1+p_2-n$, where
$n$ is an integer with $0 \le n \le N-1$,
and $0 \le z_i \le N-1$, for $i$=1,2,3.
The Clebsch-Gordan coefficients
for $V_{\epl}^{(p_i)}$ for $i=1,2,3$
are given as follows.
\bea
&& C(p_1, p_2, p_1+p_2-n; z_1, z_2, z_3)_{\epl}
= \delta(z_3, z_1+z_2-n) \non \\
&& \times {\sqrt{[2p_1+2p_2-2n+1]_{\epl}}}
{\sqrt{[n]_{\epl}! [z_1]_{\epl}!  [z_2]_{\epl}!  [z_3]_{\epl}!}}
\non \\
&& \times {\epl}^{(n-n^2)/2 + (n-z_2)p_1 +(n+z_1)p_2}
\non \\
&& \times \sum_{\nu} {\frac {(-1)^{\nu} {\epl}^{-\nu(p_1+p_2+p_3+1)} }
{[\nu]_{\epl}! [n-\nu]_{\epl}! [z_1-\nu]_{\epl}!
[z_2-n+\nu]_{\epl}!}} \non \\
&& \times {\sqrt{\frac
{[2p_1-n; z_1-\nu]_{\epl} ! [2p_1-z_1; n-\nu]_{\epl}!
 [2p_2-n; z_2+\nu -n]_{\epl} ! [2p_2-z_2; \nu]_{\epl} ! }
 {[2p_1 +2p_2 -n+1; z_1+z_2+1]_{\epl}! }}} . \non \\
\label{eq:eppp}
\eea
Here
the sum over  the integer $\nu$
in \rf{eppp} is taken
under the following condition
\be
\mbox{max}\{0, n-z_2 \}  \le \nu  \le \mbox{min}\{n, z_1 \}.
\ee

Let $V^{(j)}_{\epl}$ be the spin $j$ representation of $U_{\epl}(sl(2))$.
When $q$ is a root of unity
($q^{2N}=1$), the spin $j$ representations of $U_q(sl(2))$
has the following constraint. \cite{restur,roche}
\be
0 \le 2j \le N-2, \quad  2j \in {\bf Z}_{\ge 0}.
\ee
Let us discuss decomposition of the tensor product
\be
V^{(\mu_1)}_{\epl} \otimes V^{(\mu_2)}_{\epl}
= \sum_{p_3} N_{\mu_1 \mu_2}^{p_3} V^{(p_3)}_{\epl}
\ee
where $(\mu_1,\mu_2)=(p_1,j_2)$, $(j_1,p_2)$.
By taking the limit $q \rightarrow \epl$ in \rf{qpjp}
we have the following fusion rule.
\par \noindent
Case (1): $\mu_1=p_1, \mu_2=j_2$  \quad ($2j_2 < N-1$)
\bea
N_{p_1 j_2}^{p_3} & = & 1 \mbox{ for }
    p_3=p_1+j_2 - n , \quad 0 \le n \le 2j_2,   n \in {\bf Z} ,  \non \\
& = & 0, \quad \mbox{ otherwise} .  \label{eq:frepjp}
\eea
\par \noindent
Case (2): $\mu_1 = j_1, \mu_2=p_2$ \quad ($2j_1 < N-1$ )
\bea
N_{j_1 p_2}^{p_3} & = & 1 \mbox{ for }
    p_3=j_1+p_2 - n , \quad 0 \le n \le 2j_1,   n \in {\bf Z} ,  \non \\
& = & 0, \quad \mbox{ otherwise} .  \label{eq:frejpp}
\eea

By taking the limit: $q \rightarrow \epl$
in \rf{qpjp}, we have the Clebsch-Gordan coefficients.
Let us set $p_3=\mu_1+\mu_2-n$.
When $(\mu_1,\mu_2) = (p_1, j_2)$, we assume  that
$0 \le z_i \le N-1$ ( $i$=1,3),
$0 \le z_2 \le 2j_2$
and $0 \le n \le 2j_2$.
When $(\mu_1,\mu_2) = (j_1, p_2)$, we assume
that  $0 \le z_i \le N-1$ ( $i$=2,3),
$0 \le z_1 \le 2j_1$ and    $0 \le n \le 2j_1$.
The Clebsch-Gordan coefficients are given in the following.
\bea
&& C(\mu_1, \mu_2, \mu_1+\mu_2-n; z_1, z_2, z_3)_{\epl}
= \delta(z_3, z_1+z_2-n) \non \\
&& \times {\sqrt{[2\mu_1+2\mu_2-2n+1]_{\epl}}}
{\sqrt{[n]_{\epl}! [z_1]_{\epl}!  [z_2]_{\epl}!  [z_3]_{\epl}!}}
\non \\
&& \times {\epl}^{(n-n^2)/2 +(n-z_2)\mu_1 + (n+z_1)\mu_2}  \non \\
&& \times \sum_{\nu} {\frac {(-1)^{\nu}
{\epl}^{-\nu(2\mu_1+2\mu_2-n+1)} }
{[\nu]_{\epl}! [n-\nu]_{\epl}! [z_1-\nu]_{\epl}! [z_2-n+\nu]_{\epl}!}} \non \\
&& \times
\left( {\frac{[2\mu_1-n; z_1-\nu]_{\epl} ! [2\mu_1-z_1; n-\nu]_{\epl}!
[2\mu_2-n; z_2+\nu -n]_{\epl} ! [2\mu_2-z_2; \nu]_{\epl} ! }
{[2\mu_1 +2\mu_2 -n+1; z_1+z_2+1]_{\epl}! }} \right)^{1/2} . \non \\
\label{eq:epjp}
\eea
Here the sum over  the integer $\nu$
in \rf{epjp} is taken
under the following condition.
\par \noindent
Case (1): $\mu_1=p_1, \mu_2=j_2$
\be
\mbox{max}\{0, n-z_2 \}  \le \nu  \le \mbox{min}\{n, z_1, 2j_2-z_2 \}.
\ee
\par \noindent
Case (2): $\mu_1=j_1, \mu_2=p_2$
\be
\mbox{max}\{0, n-z_2, n+z_1-2j_1 \}  \le \nu  \le \mbox{min}\{n, z_1 \}.
\ee

The Clebsch-Gordan coefficients have the following orthogonality
relations
for the cases $(\mu_1, \mu_2)=(p_1,p_2), (p_1, j_2),
(j_1,p_2)$.

\bea
\delta_{nn^{'}}
&= & \sum_{z_1} C(\mu_1,\mu_2,\mu_1+\mu_2-n;
z_1, \tau -z_1, \tau - n)_{\epl} \non \\
& & \times C(\mu_1,\mu_2,\mu_1+\mu_2-n^{'};
z_1, \tau-z_1, \tau-n^{'})_{\epl} ,
\label{eq:oeppp1} \\
 \delta_{z_1 z_1^{'}}
&= & \sum_{n} C(\mu_1,\mu_2,\mu_1+\mu_2-n; z_1, z+n -z_1, z)_{\epl} \non \\
&& \times C(\mu_1,\mu_2,\mu_1+\mu_2-n; z_1^{'}, z+n-z_1^{'}, z)_{\epl} .
\label{eq:oeppp2}
\eea

Recall that $2\mu_i$ $(i=1,2,3)$ be either a nonnegative integer
$(2\mu_i \in {\bf Z}_{\ge 0}$) or
a complex parameter ($\mu_i \in {\bf C}$).
We introduce $\mu, \mu^{'}, \mu^{''}$ by
\be
\mu =\mu_1+\mu_2+\mu_3-n, \mu^{'}=\mu_1+\mu_2-n^{'}, \mu^{''} =
\mu_2+\mu_3-n^{''} .
\ee
where $n, n^{'}, n^{''} \in {\bf Z}_{\ge 0}$.
In terms of the Clebsch-Gordan coefficients
the Racah coefficients
$W(\mu_1,\mu_2,\mu,\mu_3;\mu^{'},\mu^{''})_{\epl}$
are given by the following.
\bea
R_{\mu^{''} \mu^{'}}({\epl}) & = & {\sqrt{[2\mu^{'}+1]_{\epl}
[2\mu^{''}+1]_{\epl}}}
W(\mu_1,\mu_2,\mu,\mu_3;\mu^{'},\mu^{''})_{\epl} \non \\
&= & \sum_{w_1}\sum_{w_2}
C(\mu_1, \mu_2, \mu^{'}; w_1, w_2, w_1+w_2-n^{'})_{\epl}
\non \\
&& \times
C(\mu^{'}, \mu_3, \mu; w_1+w_2-n^{'}, z^{''}-w_2, w_1+z^{''}-n)_{\epl} \non \\
&& \times
C(\mu_2, \mu_3, \mu_2+\mu_3-n^{''} ; w_2, z^{''}-w_2, z^{''}-n^{''})_{\epl}
\non \\
&& \times C(\mu_1, \mu^{''}, \mu; w_1, z^{''}, w_1+z^{''}+n^{''}-n )_{\epl} .
\eea

Finally we give comments. (1) The finite dimensional color representation
in this section is equivalent to the $N$ dimensional representation of
$U_q(sl(2))$ with $q^{2N}=1$ in the reference \cite{roche}.
We have discussed the representation from the view point of the
infinite dimensional representation of $U_q(sl(2))$ and the limit
$q \rightarrow \epl$ introduced in the reference \cite{inff}, and then
through the limit we have  obtained
the formulas for the Clebsch-Gordan
coefficients and the Racah coefficients.
(2) Fusion rules similar to  \rf{freppp} and \rf{frepjp}
have been given in the reference \cite{arn}. The fusion rules
are for the $m$ dimensional representations
(semi-periodic representations) of $U_q(sl(2))$ with
$q^{m}=1$, which are different from the finite dimensional
color representations (the $N$ dimensional representation with $q^{2N}=1$).

\setcounter{equation}{0}
\renewcommand{\theequation}{4.\arabic{equation}}
\section{Colored vertex models}
\subsection{Quantum affine algebra $U_q(\hat{sl(2)})$ }
We express the Boltzmann weights of the colored vertex models in terms of
the Clebsch-Gordan coefficients of color representations.
Let us consider the $q$-analog of the universal enveloping algebra
$U_q({\hat sl}(2,C))$ of
the affine Kac-Moody algebra ${\hat sl}(2)$. \cite{jimbo1}
The generators $\{X_i^{\pm}, H_i; i=0,1 \}$ satisfy the
following defining relations ($i,j=0,1)$.
\bea
\mbox{[}H_i, H_j \mbox{]} & =& 0 , \non \\
\mbox{[} H_i, X_i^{\pm} \mbox{]}  & = & \pm 2  X_i^{\pm}, \quad
\mbox{[} H_i, X_j^{\pm} \mbox{]} =  \mp 2 X_j^{\pm} \quad (i \neq j),
\non \\
\mbox{[} X_i^{+}, X_j^{-} \mbox{]} & = & \delta_{ij}
{\frac {q^{H_i}-q^{-H_i}} {q-q^{-1}} }, \non \\
\sum_{\nu=0}^{3} & (-1)^{\nu} &
{\frac {[3]_q!} {[3-\nu]_q! [\nu]_q!} }
(X^{\pm})^{3-\nu}
  X_j^{\pm} (X_i^{\pm})^{\nu} = 0  \quad (i \neq j).
\eea
The comultiplication is given by
\bea
\Delta(H_i) & = & H_i \otimes I + I \otimes H_i , \non \\
\Delta(X_i^{\pm}) & = & X_i^{\pm} \otimes q^{H_i/2}
+ q^{-H_i/2} \otimes X_i^{\pm} .
\eea
We denote by  ${\cal R}$ the universal
$R$ matrix of  ${U_q({\hat sl}(2, C))}$.
The universal $R$ matrix satisfie the following
\be
{\cal R} \Delta(a) = \tau \circ \Delta(a) {\cal R},
\quad a \in U_q({\hat sl}(2,C)),
\ee
where $\tau$ is the permutation operator
$\tau (t_1 \otimes t_2) = t_2 \otimes t_1$, for
$t_1, t_2 \in U_q({\hat sl}(2,C))$.
For simplicity we sometimes write
$U_q({\hat sl}(2,{\bf C}))$ and $U_q(sl(2,{\bf C}))$
by ${\hat U}_q$ and $U_q$, respectively.

Let us discuss solvable models from the viewpoint of $\hat{U}_q$.
We define a homomorphism $\phi$: ${\hat U}_q \rightarrow {U_q}$
by
$\phi(X_0^{\pm})  =  X^{\mp}$,
$\phi(X_1^{\pm})  =  X^{\pm}$,  and
$\phi(H_0)  =  -H$ ,  $\phi(H_1) = H$ .
We introduce
 an automorphism $T_x$ : ${\hat U}_q \rightarrow {\hat U}_q $by
$T_x (X_0^{\pm})  =  x^{\pm 1} X_0^{\pm}$,
$T_x (X)  =  X$ , for  $ X = X_1^{\pm}$ , $H_0, H_1$ .

%
We define an operator $R(x)$ by
$R(x)= \phi((T_x \otimes I ({\cal R}))$.
Then the operator $R(x)$ satisfies the following for $i=0,1$.
\bea
&& R(x) q^{{\hat H}_i/2} \otimes q^{{\hat H}_i/2}
= q^{{\hat H}_i/2} \otimes q^{{\hat H}_i/2} R(x) , \label{eq:acom} \\
&& R(x) (x^{\delta(i,0)} {\hat X}_i^{+} \otimes q^{{\hat H}_i/2}
+ q^{-{\hat H}_i/2} \otimes {\hat X}_i^{+}) \non \\
&=& (x^{\delta(i,0)} {\hat X}_i^{+} \otimes q^{-{\hat H}_i/2}
 + q^{{\hat H}_i/2} \otimes {\hat X}_i^{+}) R(x) ,
 \label{eq:bcom} \\
&& R(x) (x^{-\delta(i,0)} {\hat X}_i^{-}
\otimes q^{{\hat H}_i/2} + q^{-{\hat H}_i/2} \otimes {\hat X}_i^{-})
\non \\
& = & (x^{-\delta(i,0)} {\hat X}_i^{-}
 \otimes q^{-{\hat H}_i/2} + q^{{\hat H}_i/2}
 \otimes {\hat X}_i^{-}) R(x) .
 \label{eq:ccom}
\eea
Here
${\hat X}_i^{\pm} = \phi(X_i^{\pm})$,
${\hat H}_i = \phi(H_i)$ for $i=0,1$.

Let us take arbitrary two representations $\mu_1$, $\mu_2$ of $U_q$.
We can take as  $\mu_1$ and $\mu_2$
the spin $j$ representations of $U_q(sl(2))$,
 or the (infinite dim. or finite dim. )
color representations of $U_q(sl(2))$.
We define $R$ matrix and its matrix elements
for the representations by
 \bea
R_{\mu_1 \mu_2}(x) &= & \pi^{\mu_1} \otimes \pi^{\mu_2} (R(x)), \non \\
R_{\mu_1 \mu_2}(x)^{a_1 a_2}_{b_1 b_2}
&= & \left( \pi^{\mu_1} \otimes \pi^{\mu_2} (R(x))
\right)^{a_1 a_2}_{b_1 b_2} .\eea
Then the relations \rf{acom}, \rf{bcom} and \rf{ccom}
give linear equations
for the $R$ matrix elements.
It is easy to see
that if the matrix elements satisfy the linear equations,
then the matrix elements satisfy the Yang-Baxter equations.
\cite{jimbo0}
\bea
&& \sum_{c_1c_2c_3}
R_{p_1p_2}(u)^{c_1c_2}_{b_1b_2}
R_{p_1p_3}(u+v)^{a_1c_3}_{c_1b_3}
R_{p_2p_3}(v)^{a_2a_3}_{c_2c_3} \non \\
&=&
\sum_{c_1c_2c_3}
R_{p_1p_2}(u)^{a_1a_2}_{c_1c_2}
R_{p_1p_3}(u+v)^{c_1a_3}_{b_1c_3}
R_{p_2p_3}(v)^{c_2c_3}_{b_2b_3} .
\eea
Here we have defined the spectral parameter $u$ by $x = \exp u$.
In operator formalism, solutions of
the linear equations ($R$ matrix) can be written as
\be
R(q;u) = \sum_{\mu}
|\mu_1 \mu_2; \mu> g(\mu_1,\mu_2,\mu; u) <\mu_1 \mu_2; \mu|
\label{eq:rop}
\ee
Here
the sum is  taken over all $\mu$ appearing in
the decomposition
of the tensor product $\mu_1 \otimes \mu_2$
($\mu \subset \mu_1 \otimes \mu_2$) , and
$g(\mu_1,\mu_2,\mu; u)$ is some  function
of the spectral parameter $x=\exp u$.

\subsection{Infinite dimensional case}

Let us introduce the colored vetex models. \cite{degacts2,inff}
The Boltzmann weights of the colored vertex model
$X_{\al \beta}(u)^{ab}_{cd}$ is defined for the
configuration $\{a,b,c,d\}$. (Fig. 1)
The Yang-Baxter relation for the colored vertex model
reads
\bea
&& \sum_{c_1,c_2,c_3 } X_{\al_1,\al_2}( u_1-u_2)^{a_1 a_2}_{c_2 c_1}
X_{\al_1,\al_3}( u_1-u_3)^{c_1 a_3}_{c_3 b_1}
X_{\al_2,\al_3}(u_2-u_3)^{c_2 c_3}_{b_3 b_2} \nonumber \\
&= & \sum_{c_1,c_2,c_3}
X_{\al_2,\al_3}(u_2-u_3)^{a_2 a_3}_{c_3 c_2}
X_{\al_1,\al_3}(u_1-u_3)^{a_1 c_3}_{b_3 c1}
X_{\al_1,\al_2}(u_1-u_2)^{c_1 c_2}_{b_2 b1} .
\nonumber \\
\label{cybr}
\eea

We construct the colored vertex models
by calculating the matrix elements of the
$R$ matrix \rf{rop} on the color representations.
Let us consider color representations
$(\pi^{(p_i)}_{q}, V^{(p)}) \quad (i=1,2)$
of $U_q(sl(2))$.
We define $R_{p_1p_2}(q; u)$ and its matrix elements by
\bea
R_{p_1p_2}(q; u) & = & \pi^{(p_1)}_q \otimes \pi^{(p_2)}_q (R(x)),
\non \\
R_{p_1p_2}(q; u)^{a_1 a_2}_{b_1 b_2} & = &
\left( \pi^{(p_1)}_q \otimes \pi^{(p_2)}_q (R(x))
\right)^{a_1 a_2}_{b_1 b_2} .
\eea
Here $a_1, a_2, b_1, b_2 =0, 1, \cdots, \infty$.

We can show that
the follwoing gives a solution of
the linear equations \rf{acom}, \rf{bcom} and \rf{ccom}.
\bea
R_{p_1p_2}(q; u)^{a_1 a_2}_{b_1 b_2}
& = & \sum_{n}^{\infty} g(p_1,p_2,n;u)_q  \non \\
&& C(p_1,p_2,p_1+p_2-n;b_1,b_2, b_1+b_2-n)_q  \non \\
&& C(p_2,p_1,p_1+p_2-n;a_2,a_1, a_1+a_2-n)_q  . \label{eq:ipp}
\eea
where
\be
g(p_1,p_2,n; u)_q = (-1)^n \prod_{k=0}^{n-1}
{\frac {[u-p_1-p_2+k]_q} {[u+p_1+p_2-k]_q} },
\mbox{ for } n \in {\bf Z}_{\ge 0} .  \label{eq:qg}
\ee
We assume $g(p_1, p_2,n=0; u)_q=1$.
For  the fixed values of $a_i,b_i (i=1,2)$,  the
sum in \rf{ipp} reduces to a finite sum. Therefore the sum
in \rf{ipp}   is  well defined.
Thus we have an explicit formula for the $R$ matrix elements for
the infinite dimensional color representation.
It is remarked that the case $p_1=p_2$, the
infinite dimensional
representation of the $R$ matrix with a spectral parameter
were discussed by Jimbo by using projection operators.
\cite{jimbo1}

Through the $R$ matrix \rf{ipp}
the Boltzmann weights of the colored vertex model
are written as follows
\be
X_{\al \beta}^{(\infty)}(u)^{ab}_{cd} = R_{p_1 p_2}(q;u)^{ba}_{cd} ,
\ee
where $\al = q^{p_2}$ and $\beta = q^{p_1}$.
Thus we obtain
the colored vertex model of the infinite
state case introduced in the reference \cite{inff}.

Using the Clebsch-Gordan coeffcients we can calculate
$R$ matrix for the representations $V^{j}$ and $V^{(\infty)}$.
We define $R_{p_1j_2}(x)$ and its matrix elements by
\bea
R_{p_1j_2}(x) & = & \pi^{(p_1)}_q \otimes \pi^{(j_2)}_q (R(x)),
\non \\
R_{p_1j_2}(x)^{a_1a_2}_{b_1b_2} & = &
\left( \pi^{(p_1)}_q \otimes \pi^{(j_2)}_q (R(x))
\right)^{ab}_{cd} .
\eea
Here $a_1,b_1 =0, 1, \cdots, N-1$, and $a_2, b_2 = 0, 1, \cdots 2j_2$.
The matrix elements are given by the following.
\bea
 R_{p_1j_2}(x)^{a_1 a_2}_{b_1 b_2}
& = & \sum_{n=0}^{2j_2} g(p_1,j_2,n;u)_q  \non \\
&& C(p_1,j_2,p_1+p_2-n;b_1,b_2,b_1+b_2-n)_q  \non \\
&& C(j_2,p_1,p_1+p_2-n;a_2,a_1,a_1+a_2-n)_q  . \label{eq:ipj}
\eea
Here $g(p_1,j_2,n;u)_q$ is given by \rf{qg} with $p_2=j_2$.
%
%

\subsection{Finite dimensional case}

Let us consider color representations
$(\pi^{(p_i)}_{\epl}, V^{(N)}) \quad (i=1,2)$
of $U_q(sl(2))$ with $q=\epl$.
We define $R_{p_1 p_2}(\epl; x)$ and its matrix elements by
\bea
R_{p_1 p_2}(\epl; x) & = & \pi^{(p_1)}_{\epl}
\otimes \pi^{(p_2)}_{\epl} (R(x)),
\non \\
R_{p_1 p_2}(\epl; x)^{a_1 a_2}_{b_1 b_2} & = &
\left( \pi^{(p_1)}_{\epl} \otimes \pi^{(p_2)}_{\epl} (R(x))
 \right)^{a_1 a_2}_{b_1 b_2}.
\eea
Then we have the following.
\bea
 R_{p_1p_2}(\epl; u)^{a_1 a_2}_{b_1 b_2}
& = & \sum_{n=0}^{N-1}  g(p_1,p_2,n; u)_{\epl}  \non \\
&& \times  C(p_1,p_2,p_1+p_2-n;b_1,b_2,b_1+b_2-n)_{\epl}  \non \\
&& \quad  \times  C(p_2,p_1,p_1+p_2-n;a_2,a_1,a_1+a_2-n)_{\epl},
\label{eq:fpp}
\eea
where
\be
g(p_1,p_2,n; u)_{\epl} = (-1)^n \prod_{k=0}^{n-1}
{\frac {[u-p_1-p_2+k]_{\epl}} {[u+p_1+p_2-k]_{\epl}} },
\mbox{ for } n \in {\bf Z}_{\ge 0} .  \label{eq:eg}
\ee
We assume $g(p_1,p_2,n=0; u)_{\epl}=1$.

Through the $R$ matrix \rf{ipp}
the Boltzmann weights weights of the colored vertex model
are written as follows
\be
X_{\al \beta}^{(N)}(u)^{ab}_{cd} = R_{p_1 p_2}(\epl; u)^{ba}_{cd} ,
\ee
where $\al = {\epl}^{p_2}$ and $\beta = {\epl}^{p_1}$.
Thus we obtain
the colored vertex model of the finite
state case given in the reference \cite{degacts2}.

Using the Clebsch-Gordan coeffcients we can calculate
$R$ matrix for the representations $V^{j}$ and $V^{(N)}$.
We define $R_{p_1j_2}(x)$ and its matrix elements by
\bea
R_{p_1j_2}(\epl; x) & = & \pi^{(p_1)}_{\epl}
\otimes \pi^{(j_2)}_{\epl} (R(x)),
\non \\
R_{p_1j_2}(\epl; x)^{a_1a_2}_{b_1b_2} & = &
\left( \pi^{(p_1)}_q \otimes \pi^{(j_2)}_{\epl} (R(x))
\right)^{ab}_{cd} .
\eea
Here $a_1,b_1 =0, 1, \cdots, N-1$, and $a_2, b_2 = 0, 1, \cdots 2j_2$.
The matrix elements are given by the following.
\bea
 R_{p_1j_2}(\epl; x)^{a_1 a_2}_{b_1 b_2}
& = & \sum_{n}^{2j_2} g(p_1,j_2,n;u)_{\epl}   \non \\
&& C(p_1,j_2,p_1+p_2-n;b_1,b_2,b_1+b_2-n)_{\epl}  \non \\
&& C(j_2,p_1,p_1+p_2-n;a_2,a_1,a_1+a_2-n)_{\epl}  . \label{eq:fpj}
\eea
Here the function $g(p_1,j_2,n; u)_{\epl}$ is given by
\rf{eg} with $p_2=j_2$.
It is easy to show  that
the $R$ matrix \rf{fpj}  gives a solution of
the linear equation \rf{bcom}.

\setcounter{equation}{0}
\renewcommand{\theequation}{5.\arabic{equation}}
\section{Colored IRF models}
\subsection{Infinite dimensional case}
Using the Racah coefficients for color representations
we define colored  IRF (Interaction Round a Face) models.
Let the symbol
$w(a, b, c, d ;p_1,p_2 ; u)_q$
denotes the Boltzmann weight of
colored IRF model for the configuration
$\{ a, b, c, d ;p_1,p_2 \}$.  (Fig. 2)
The admissible condition is determined by the fusion rules.
In the configuration
$\{ a, b, c, d ;p_1,p_2 \}$,
$a$ is admissble to $d$ if $a$ $\subset  p_1 \otimes d $, i.e.,
$N_{p_1, d}^{a} \neq 0$.  We denote this condition by $a \sim d$.
The Boltzmann weight
$w(a, b, c, d ;p_1,p_2 ; u)_q$
is defined  to be zero,
unless $a \sim d$, $b \sim a$ , $c \sim d$ and $b \sim c$.

The Yang-Baxter relation for the
colored IRF model is given by the following.
\bea
&& \sum_{g} w(g_1,g_2,g,g_0; p_1,p_2;u) w(g_2,g_3,g_2^{'},g; p_1,p_3; u+v)
\non \\
&& w(g,g_2^{'},g_1^{'},g_0; p_2,p_3; v) \non \\
&= &\sum_{g} w(g_2,g_3,g,g_1;p_2,p_3;v)w(g_1,g,g_1^{'},g_0;p_1,p_3;u+v)
\non \\
&& w(g,g_3,g_2^{'},g_1^{'};p_1,p_2;u) . \label{eq:qirfybr}
\eea

The Boltzmann weights of the colored IRF models are given as follows.
\bea
&&  w( g_{12}, g_1, g_{21}, g_0 ;p_1, p_2; u)_q \non \\
& = & \sum_{n=0}^{\infty}
g(p_1, p_2, n; u)_q
[2p+1]_q {\sqrt{[2g_{12}+1]_q[2g_{21}+1]_q}} \non \\
&& W(p_2, p_1, g_1, g_0 ; p, g_{12})_q
W(p_2, g_1, p_1, g_0 ; g_{21}, p)_q , \label{eq:qirfpp}
\eea
where $p=p_1+p_2-n, n \in {\bf Z}_{\ge 0}$, and the
function
$g(p_1, p_2, n; u)_q$ is given by \rf{qg} .
We recall that the symbols
$W$ and $w$ denote the Racah coefficient and the Boltzmann weight of
the IRF model, respectively.

The expression \rf{qirfpp} can be derived from \rf{ipp}
by using the definition of the Racah coefficients \cite{pasq}
(see also \cite{nomura2} for the cases of the spin $j$ representations).
Therefore the Boltzmann weights given in
\rf{qirfpp}
satisfy the Yang-Baxter relation for the IRF models \rf{qirfybr}.

We can discuss hybrid type colored IRF models, i.e.,
$2\mu_1 \in {\bf Z}_{\ge 0}$ or
$2\mu_2 \in {\bf Z}_{\ge 0}$. For simplicity we consider only the case
$p_1 \in {\bf C}$ and $2\mu_2=2j_2 \in {\bf Z}_{\ge 0}$.
The Boltzmann weights of the colored IRF models are given as follows.
\bea
&& w(g_{12}, g_1, g_{21}, g_0 ; p_1, j_2; u)_q \non \\
& = & \sum_{n=0}^{2j_2}
g(p_1, j_2, n; u)_q
[2p+1]_q {\sqrt{[2g_{12}+1]_q[2g_{21}+1]_q}} \non \\
&& W(j_2, p_1, g_1, g_0 ; p, g_{12})_q
W(p_1, j_2, g_1, g_0; g_{21}, p)_q ,
\eea
where $p=p_1+j_2-n, 0 \le n \le 2j_2,
n \in {\bf Z}_{\ge 0}$, and the function
$g(p_1, p_2, n; u)_q$ is given by \rf{qg}.

\subsection{Finite dimensional case}

Using the Racah coefficients for color representations
we define colored  IRF (Interaction Round a Face) models.
Let the symbol
$w(a, b, c, d ;p_1,p_2 ; u)_{\epl}$
denote the Boltzmann weight of
colored IRF model associated with finite dimensional
color representations of $U_q(sl(2))$.

The Boltzmann weights of the colored IRF model are given by the following.
\bea
&& w( g_{12}, g_1, g_{21}, g_0 ;p_1, p_2;u)_{\epl} \non \\
& = & \sum_{n=0}^{N-1}
g(p_1, p_2, n; u)_{\epl}
[2p+1]_q {\sqrt{[2g_{12}+1]_{\epl}[2g_{21}+1]_{\epl}}}
\non \\
&& \times W(p_2, p_1, g_1, g_0 ; p, g_{12})_{\epl}
W(p_1, p_2, g_1, g_0 ; p, g_{21})_{\epl}
\eea
Here the function $g(p_1, p_2, n; u)_{\epl}$ is given by \rf{eg}.

We can discuss hybrid type colored IRF models, i.e.,
$2\mu_1 \in {\bf Z}_{\ge 0}$ or
$2\mu_2 \in {\bf Z}_{\ge 0}$. For simplicity we consider only the case
$p_1 \in {\bf C}$ and $2\mu_2=2j_2 \in {\bf Z}_{\ge 0}$.
The Boltzmann weights of the colored IRF models are given as follows.
\bea
&& w(g_{12}, g_1, g_{21}, g_0 ; p_1, j_2; u)_{\epl} \non \\
& = & \sum_{n=0}^{2j_2}
g(p_1, j_2, n; u)_{\epl}
[2p+1]_{\epl} {\sqrt{[2g_{12}+1]_{\epl} [2g_{21}+1]_{\epl}}} \non \\
&& W(j_2, p_1, g_1, g_0 ; p, g_{12} )_{\epl}
W(p_1, j_2, g_1, g_0 ; p, g_{21})_{\epl}
\eea
where $p=p_1+j_2-n$, $
0 \le n \le 2j_2, n \in {\bf Z}_{\ge 0}$, and the function
$g(p_1, j_2, n)_{\epl}$ is given by \rf{eg}.

\setcounter{equation}{0}
\renewcommand{\theequation}{6.\arabic{equation}}
\section{Invariants of trivalent colored oriented graphs}
\subsection{Basic relations for the colored braid matrices}
We construct new  invariants of colored oriented graphs
using the Clebsch-Gordan coefficients for finite dimensional
color representations given in \S 3.

Let us define the $N$-state colored braid matrix.
We assume the charge conservation:
$G^{ab}_{cd}(p_{\al}, p_{\beta}; \pm) =0$ unless $a+b=c+d$
($\alpha=q^{-4p_{\al}}$, $\beta=q^{-4p_{\beta}}$).
Furthermore we assume
$G^{ab}_{cd}(p_{\al}, p_{\beta}; +) = 0$ if $a< d$, and
$G^{ab}_{cd}(p_{\al}, p_{\beta}; -) = 0$ if $a> d$.
We define $(z;q)_n$ by
$(z;q)_n = (1-z)(1-zq)\cdots(1-zq^{n-1})$ for $n>0$ and
$(z;q)_0=1$ for $n=0$.
The nonzero elements are written as follows.
\bea
&& G^{a,b}_{c,d}(p_{\alpha},p_{\beta}; +) =
{\epl}^{-2bp_{\al}-2dp_{\beta}}
{\epl}^{2bd}{\epl}^{2p_{\alpha}p_{\beta}}
\non \\
&&
\left(
{\frac {({\epl}^{2};{\epl}^{2})_{c}}
 {({\epl}^{2};{\epl}^{2})_{b}
({\epl}^{2};{\epl}^{2})_{c-b} } }
{\frac {({\epl}^{2};{\epl}^{2})_{a}}
 {({\epl}^{2};{\epl}^{2})_{d}
({\epl}^{2};{\epl}^{2})_{a-d} } }
{\frac {({\epl}^{-4p_{\alpha}};{\epl}^{2})_{a}}
{({\epl}^{-4p_{\al}};{\epl}^{-2})_{d}} }
{\frac {({\epl}^{-4p_{\beta}};{\epl}^{2})_{c}}
{({\epl}^{-4p_{\beta}};{\epl}^{2})_{b}} } \right)^{1/2}, \non \\
&& G^{a,b}_{c,d}(p_{\al},p_{\beta}; -) =
{\epl}^{2ap_{\al}+2cp_{\beta}}
{\epl}^{-2ac}{\epl}^{-2p_{\alpha}p_{\beta}}
\non \\
&& \left( {\frac {({\epl}^{-2};{\epl}^{-2})_{b}}
 {({\epl}^{-2};{\epl}^{-2})_{b-c}
({\epl}^{-2};{\epl}^{-2})_{c} } }
{\frac {({\epl}^{-2};{\epl}^{-2})_{d}}
 {({\epl}^{-2};{\epl}^{-2})_{d-a}
({\epl}^{-2};{\epl}^{-2})_{a} } }
{\frac {({\epl}^{4p_{\al}};{\epl}^{-2})_{b}}
{({\epl}^{4p_{\al}};{\epl}^{-2})_{c}} }
{\frac {({\epl}^{4p_{\beta}};{\epl}^{-2})_{d}}
{({\epl}^{4p_{\beta}};{\epl}^{-2})_{a}} } \right)^{1/2} . \non \\
\eea
Then the colored braid matrix satisfies the following relation.
\begin{eqnarray}
& & \sum_{c_1,c_2,c_3}
G^{a_1,a_2}_{c_2,c_1}(p_1,p_2; +)
G^{c_1,a_3}_{c_3,b_1}(p_1,p_3; +)
G^{c_2,c_3}_{b_3,b_2}(p_2,p_3; +)
\nonumber \\
& = &
 \sum_{c_1,c_2,c_3}
G^{a_2,a_3}_{c_3,c_2}(p_2,p_3; +)
G^{a_1,c_3}_{b_3,c_1}(p_1,p_3; +)
G^{c_1,c_2}_{b_2,b_1}(p_1,p_2; +) .  \label{eq:cbr}
\end{eqnarray}
Here $p_1,p_2, p_3$ denote the colors of the strings.

%
%
It has been explicitly shown that the colored braid matrix is equivalent to
the $R$ matrix of the color representations of $U_q(sl(2))$.
\cite{inff}

Let us introduce basic relations for the colored braid matrix and the
Clebsch-Gordan coefficients. We note they are related to the
Reidemeister moves.
\par \noindent
(i) First inversion relation:
\begin{equation}
\sum_{e,f}
G^{ab}_{ef}(p_1,p_2;  +)
G^{ef}_{cd}(p_1,p_2; -)
=\sum_{e,f}
G^{ab}_{ef}(p_1,p_2; -)
G^{ef}_{cd}(p_1,p_2; +)
=\delta^a_c \delta^b_d . \label{eq:br1}
\end{equation}
\par \noindent
(ii) Second inversion relation:
\begin{eqnarray}
\sum_{e,f}G^{ae}_{cf}(p_1,p_2; +)G^{df}_{be}(p_1,p_2; -)
      \om^{{  d} -{  e}}&=& \delta^a_b \delta^c_d ,
			\non \\
\sum_{e,f}G^{ae}_{cf}(p_2, p_1; -)G^{df}_{be}(p_2, p_1; +)
      \om^{{  d} -{  e}}&=& \delta^a_b \delta^c_d .  \non \\
&& \label{eq:br2}
\end{eqnarray}
\par \noindent
(iii) Markov trace property:
\be
\sum_b G^{ab}_{ab}(p,p; \pm)\om^{-b}={\epl}^{2p(N-1) \pm 2p(N-1) \pm 2p^2} .
\label{eq:br3}
\ee
%
%
\par \noindent
(iv)
\bea
&&\sum_{c_1c_2} G^{a_1 c_3}_{b_3 c_1}(p_1, p_3;+)
G^{a_2 a_3}_{c_3 c_2}(p_2, p_3; +)
C(p_1,p_2, p; c_1, c_2, b)_{\epl}  \non \\
& = & \sum_{c}
C(p_1,p_2,p; a_1, a_2, c)_{\epl} G^{c a_3}_{b_3 b}(p, p_3; +),   \non \\
&&\sum_{c_1c_2} G^{a_1 c_3}_{b_3 c_1}(p_3, p_1;-)
G^{a_2 a_3}_{c_3 c_2}(p_3, p_2; -)
C(p_1,p_2, p; c_1, c_2, b)_{\epl}  \non \\
& = & \sum_{c}
C(p_1,p_2,p; a_1, a_2, c)_{\epl} G^{c a_3}_{b_3 b}(p_3, p; -) .
\label{eq:br4}
\eea

%
%

\subsection{Invariants of colored oriented  framed
tangle graphs
with trivalent vertices}

We consider trivalent graphs, which have
vertices with three edges.
Framed graphs have framing vector fields.
We assume that all three edges have common tangent vector
in the trivalent vertex, and the framing vector field is always normal
to the tangent vector to the vertex. (Fig. 3)

We introduce "tangle graph".
We define  $(k,l)$-oriented tangle graph $T$ by
a finite set of disjoint oriented arcs,
oriented trivalent vertices, and oriented circles
properly embedded
in ${\bf R}^2 \times [0,1]$ such that
\be
\partial T = \{(i,0,0); i=1,2, \cdots , k \} \cup
\{(j,0,1); j=1,\cdots, l  \},
\ee
where $\partial T$ denotes the upper and lower boundaries of
the tangle graph. (Fig. 4)
We define colored oriented tangle graphs $(T, \Al)$ by
 assigning  colors on edges, arcs and circles
 of tangle graphs.

We can express
trivalent framed tangle graph by plane diagram.
\cite{turaev,kau,piu}
We can choose the diagram such that the framing
vector field is normal to the plane of projection and directed "up".
\begin{prop} \cite{turaev}
The isotopy invariants of framed tangle graphs are
 functions on their plane
diagrams, invariant under the following local moves
$D_1 \sim D_7$.
\end{prop}
(Fig. 5)

We introduce weights for diagrams.
We assign the colored braid matrices, the Clebsch-Gordan coefficients,
and the weights $U_i$ to the braiding diagrams, trivalent vertex diagrams,
and the creation-annihilation diagrams, respectively. (Fig. 6)

We define  $\phi(T, \Al)$ for a tangle graph $T$ by the  summation over
all possible configurations of variables $z_i$ on the edges (or
segments) of the graph. In the summation we fix the colors $p_i$.
The sum corresponds to partition function in statistical mechanics.
Then from the relations \rf{cbr}, \rf{br1}, \rf{br2}, \rf{br3}  the sum
$\phi(T, \Al)$
for the tangle graph $T$ is
invariant under the moves
$D_1 \sim D_7$.
Thus we have isotopy invariants of trivalent
colored oriented framed tangle graphs.

\subsection{Invariants of trivalent framed graphs}

We construct invariants of trivalent
colored oriented framed  graphs by
an approach parallel to that for the colored link invariants. \cite{ado}
Through $(1,1)$-tangle graph
we introduce another invariant of a framed graph. \cite{ado}

Let $T$ be a $(1,1)$- tangle graph.
We denote by $\hat T$ the graph obtained
by closing the open strings of $T$.
It is easy to show the following proposition. \cite{ado}
\begin{prop}
Let $T_1$ and $T_2$ denote two $(1,1)$-tangle graph.
If $\hat T_1$ is isotopic to $\hat T_2$
as a graph in $S^3$ by an isotopy
which carries the closing component of $\hat T_1$
to that of $\hat T_2$.
Then $T_1$ is isotopic to $T_2$ as a $(1,1)$-tangle graph.
\end{prop}

Let $T$ be a $(1,1)$-tangle graph.
We put $F=\hat T$
and $s$ is the color of the
closing component (or edge) of $\hat T$.
We denote by
$\phi(T,\Al)^{a}_b$
the value $\phi$ for the tangle graph
with variables
$a$ and $b$
on the closing component (or edge). (Fig. 7)

Then from the discussion given in the reference
\cite{ado}
we can show that
\be
\phi(T,\Al)^a_b = \lm \delta_{ab} .
\ee
The value of  $\phi(T, \Al)^{a}_b$ do not depend on
$a$ or $b$.

For a colored graph $(F,\Al)$ and a color
$s$ of closing component (or edge),
we define $\Phi$ by $\Phi(F,s,\Al) = \lm$
where $F,T,s$ are above  and
$\phi(T,\Al)^a_b = \lm \delta_{ab}$.
By the basic relations given in \S 6.2,
$\Phi$ is well-defined, i.e. $\Phi(F,s,\Al)$
does not depend on a choice of $T$.

Further we have the next proposition, to obtain invariants
which do not depend on $s$.
\begin{prop} \cite{ado}
For a graph $F$ and its color $\Al = (p_1, \cdots, p_n)$,
we have the next formula.
\be
 \Phi(F,s,\Al)
([p_s; N-1]_{\epl}!)^{-1}
=\Phi(F,s',\Al)
([p_{s'}; N-1]_{\epl}!)^{-1} .
\ee
\end{prop}
The proof of
this proposition
is equivalent to that given in Appendix C of the reference \cite{ado}.
By this proposition we obtain the next definition.

\begin{df}
For a trivalent colored oriented framed graph $(F,\Al)$,
we define an isotopy invariant $\hat \Phi $ of $(F,\Al)$ by
\be
\hat \Phi(F,\Al) = \Phi (F,s,\Al)
([p_s; N-1]_{\epl}!)^{-1} .
\ee
\end{df}
Thus we obtain new invariants
$\Phi(F, \Al)$
of trivalent
colored oriented framed graphs  $(F, \Al)$.

{\hskip 2.0cm}
\par \noindent
{\bf Acknowledgement}

We would like  to thank Prof. M. Wadati for encouragement.

\newpage

\setcounter{equation}{0}
\renewcommand{\theequation}{A.\arabic{equation}}
\appendix
\section{Appendix A}

We give derivation of the Clebsch-Gordan coefficients
for infinite dimensional representations of $U_q(sl(2))$.
We calculate the Clebsch-Gordan coefficients following
Racah's approach \cite{rose}.
For simplicity we  denote $C(p_1,p_2,p_1+p_2-n; z_1,z_2,z)_q$
by $C(n; z_1,z_2,z)$.

Applying  $\Delta(X^{\pm})$ on $|p_1,p_2; p,z>$
($p=p_1+p_2-n$), we have
\bea
&& \sqrt{[z][2p+1-z]}C(n; z_1,z_2-1,z-1) \non \\
&=& \sqrt{[z_1+1][2p_1-z_1]}q^{p_2-z_2+1} C(n;z_1+1,z_2-1,z) \non \\
&& +q^{-(p_1-z_1)}\sqrt{[z_2][2p_2+1-z_2]}C(n;z_1,z_2,z), \\
&& \sqrt{[z+1][2p-z]} C(n; z_1,z_2+1,z+1) \non \\
&=& \sqrt{[z_1][2p_1+1-z_1]}q^{p_2-z_2-1} C(n;z_1-1,z_2+1,z) \non \\
&& +q^{-(p_1-z_1)}\sqrt{[z_2+1][2p_2-z_2]}C(n;z_1,z_2,z).
\eea
We introduce $f(n; z_1,z_2,z)$ by
\bea
&& C(n;z_1,z_2,z)=(-1)^{z_1}q^{(p_1-z_1)(p-z+1)} \non \\
&& \times \left( {\frac {[2p_1-z_1; n-z_1]! [2p_2-z_2;n-z_2]!}
{[2p; z]![z_1]![z_2]![z]! } } \right)^{1/2}
f(n; z_1,z_2,z).
\eea
Then we have the following recurrence relations.
\bea
& & f(n;z_1,z_2+1) = -[z_1][2p_1+1-z_1]q^{2p-2z}f(n;z_1-1,z_2+1,z) \non \\
 & & \quad + [2p_2-z_2][z_2+1]f(n; z_1,z_2), \label{eq:rc1} \\
& & [z][2p-z+1] q^{2(p_1-z_1)} f(n;z_1,z_2-1)  \non \\
&& = -f(n;z_1+1,z_2-1,z) +f(n; z_1,z_2,z). \label{eq:rc2}
\eea
Setting $z=0$ in \rf{rc2} we have  $f(n;z_1,z_2,0)=f(n;z_1+1,z_2-1,0)$.
Therefore we write $f(n; z_1, z_2, 0)$ as $f_n$. From \rf{rc1} we have
\bea
f(n;z_1,z_2,z)& =& f_n [z_1]! [z_2]!  \sum_t (-1)^t q^{t(2p+1-z)} \qc{z}{t}
\non \\
&& \times {\frac {[2p_1-z_1+t;t]! [2p_2-z_2+z-t; z-t]!}
{[z_1-t]! [z_2-z+t]!} } . \label{eq:f}
\eea
By using \rf{fm2} it is easy to show that \rf{f} satisfies
the recurrence relation \rf{rc1}.

\par
Now we consider $f_n$.
If we assume the following expression of $C(n;z_1,z_2,0)$
\bea
C(n; z_1,z_2,0) &=& (-1)^{z_1}q^{n(1-n)/2-z_1(p+1) +np_2}  \non \\
&& \times \left( {\frac {[n]! [2p_1-z_1;n-z_1]! [2p_2-z_2;n-z_2]! }
{[2p+n+1; n]! [z_1]! [z_2]!}} \right)^{1/2}, \label{eq:c0}
\eea
we can  show the orthogonality relation
\be
\sum_{z_1} C(m; z_1,n-z_1,n-m) C(n;z_1,n-z_1,0) = \delta_{nm},
\quad 0 \le m \le n.
\label{eq:ort}
\ee
We prove \rf{ort} in the following.
When  $m=n$,
we show \rf{ort}
using  \rf{fm5} where $\al=2p_1-n$ and $\beta=2p_2-n$.
When $m < n$, we prove \rf{ort}
  putting the expression \rf{f} and \rf{c0} in
 the relation \rf{ort},
 transforming the variables $\{z_1,t\}$ into
$\{t, \nu \}$ ($z_1-t=\nu$) and then taking the sum over $t$ using \rf{fm3}.
Then we see that the sum vanishes.

{}From the relation \rf{ort} and the convention of phase factor,
 the expression \rf{c0} is shown.
 We obtain $f_n$ as follows.
\be
f_n = q^{n(1-n)/2+np_2-p_1(p+1)}
\sqrt{\frac {[n]!} {[2p+n+1;n]!} } .
\ee
Thus we have the following expression for the Clebsch-Gordan coefficients.
\bea
C(n;z_1,z_2,z)& =& \delta(z_1+z_2-n,z)q^{n(1-n)/2+np_2
-z_1(p+1) - (p_1-z_1)z}  \non \\
&& \times \left( {\frac {[2p_1-z_1; n-z_1]! [2p_2-z_2;n-z_2]!
[n]![z_1]![z_2]![z]!}
{[2p+n+1;n]![2p;z]!} } \right)^{1/2} \non \\
&& \sum_t (-1)^{t+z_1} q^{t(2p+1-z)}
{\frac    {[2p_1-z_1+t; t]! [2p_2-z_2+z-t; z-t]!}
{[t]![z-t]![z_1-t]![z_2-z+t]!} } . \non \\
\label{eq:racah}
\eea

Finally we transform the expression \rf{racah} of
the Clebsch-Gordan coefficients into the symmetric one
\rf{qppp} given in the section 3.
We apply the formula \rf{fm4} to the expression \rf{racah}
( $\al=2p_1-z_1, \beta=t, c=n-z_1+t$), and
replace the variable $k$  by $k=t-u$. We apply  the formula \rf{fm6}
($\al=2p_2+z-z_2-u, s=t-u, a=z_1-u, b=z-u$). Then we
replace the variable $u$ by $\nu=z_1-u$. We thus obtain the symmetric
expression \rf{qppp}.

We have considered only the case $p_i \in {\bf C}$. The Clebsch-Gordan
coefficients for the other cases
such as $(\mu_1,\mu_2) = (p_1,j_2)$ can be derived in the same way.

\setcounter{equation}{0}
\renewcommand{\theequation}{B.\arabic{equation}}
\section{Appendix B}

We define $q$-analog of combinatorial for $m \ge k$,
$m,k \in {\bf Z}_{\ge 0}$ and $\alpha \in {\bf C}$
as follows.
\be
\qc{m}{k} = {\frac {[m]!} {[m-k]![k]!}}, \quad
\qc{\al}{k} = {\frac {[\al;k]!}{[k]!}}.
\ee

\par \noindent (1)
For  $m,n \in {\bf Z}_{\ge 0}$ ($n \ge m$) ,
we have the following useful formulas.
\bea
&& \prod_{k=1}^m (1-zq^{2k-2}) = \sum_{k} (-z)^k q^{k(m-1)} \qc{m}{k},
\label{eq:fm1} \\
&& \qc{n-1}{m} + \qc{n-1}{m-1}q^{\pm n} = \qc{n}{m}q^{\pm m},
\label{eq:fm2} \\
&& \sum_{k=0}^{n-m} (-1)^k \qc{n-m}{k}q^{k(n-m-1)} = \delta_{mn} .
\label{eq:fm3}
\eea

\par \noindent (2)
For $\al, \beta \in {\bf C}$, and $m, n, a, b, c  \in {\bf Z}_{\ge 0}$,
we have the following.
\bea
&& \sum_{k} \qc{\al}{c-k} \qc{\beta}{k}q^{\pm k(\al+\beta)}
= \qc{\al+\beta}{c} q^{\pm c \beta},
\label{eq:fm4} \\
&& \sum_k \qc{\al+n-k}{n-k} \qc{\beta+k}{k} q^{-k(\al+\beta+2)}
= \qc{\al+\beta+n+1}{n}q^{-n(1+\beta)}, \label{eq:fm5} \\
&& \sum_s (-1)^s \qc{\al-s}{b-s} \qc{a}{s} q^{-s(-\al+a+b-1)}
= \qc{\al-a}{b} q^{-ab}.  \label{eq:fm6}
\eea

We can derive \rf{fm1} induction on $m$.
We derive the relation \rf{fm2} from \rf{fm1}.
The relation \rf{fm3} follows  \rf{fm1}.

We show the relation \rf{fm4} first for the case $\al=m_1, \beta=m_2$
($m_1, m_2 \in {\bf Z}_{\ge 0}$)
by using \rf{fm1}.  Since $m_1, m_2$ are arbitrary
positive integers, and \rf{fm4} is equivalent to
a polynomial relation in terms of $q^{\al}$ and $q^{\beta}$,
the relation \rf{fm4} holds also
for complex parameters $\alpha, \beta$.
{}From \rf{fm4}
we derive the relations \rf{fm5} and \rf{fm6} using the following
\be
\qc{\al}{n} = (-1)^n \qc{-\al-1+n}{n}, \quad \mbox{ for }
\al \in {\bf C}, n \in {\bf Z}_{\ge 0}.
\ee

\newpage
\par \noindent
{\bf Figure Captions }
\par \noindent
Fig. 1
\par
(a)  Boltzmann weight $X_{\alpha \beta}(u)^{ab}_{cd}$ and
(b)  $R$ matrix $R_{p_1 p_2}(u)^{a_1 a_2}_{b_1 b_2}$
\par \noindent
Fig. 2
\par
Boltzmann weight $w(a,b,c,d; p_1, p_2; u)$
\par \noindent
Fig. 3
\par
Trivalent vertex. The edges have colors $p_1, p_2, p$ and
variables $z_1, z_2, z$

\par \noindent
Fig. 4
\par
(2,3)-tangle graph. Each segment in the graph has
a color $p_j$.

\par \noindent
Fig. 5
\par
Relations D1 $\sim$ D7.

\par \noindent
Fig. 6
\par
We assign the follwoing weights to the diagrams.
\par
(a) Identity diagram. $I^a_b = \delta_{ab}$,
 $(I^*)^a_b  = \delta_{ab}$
\par
(b) Creation-annihilation diagrams. \par
$(U_r)_{ab} = q^{-p(N-1)}{\epsilon}^{-b} \delta_{a+b, N-1}$ \par
$(U_l)_{ab} = q^{p(N-1)}{\epsilon}^a \delta_{a+b, N-1}$ \par
$({\bar U}_r)_{ab} = q^{p(N-1)}{\epsilon}^{a} \delta_{a+b, N-1}$ \par
$({\bar U}_l)_{ab} = q^{-p(N-1)}{\epsilon}^{-b} \delta_{a+b, N-1}$ \par
\par
(c) Braiding diagrams
\par
$G(p_1,p_2; +)^{ab}_{cd}$
and
$G(p_1,p_2; -)^{ab}_{cd}$
\par
(d) Vertex diagrams.
\par
We assign
the Clebsch-Gordan coefficient
$C_(p_1,p_2,p; z_1,z_2,z)$.
both to the two vertex diagrams $V$ and $V^{'}$.

\par \noindent
Fig. 7
\par
The color $s$ of the closing edge (or component) is
 $p_3$.

\end{document}